\documentclass[preprint2]{aastex}

\newcommand{\bleo}{$\beta$~Leo}
\newcommand{\dleo}{$\delta$~Leo}
\newcommand{\zlep}{$\zeta$~Lep}

\slugcomment{Accepted for publication in the Astrophysical Journal}

\shorttitle{Inner regions of debris disks}
\shortauthors{Akeson et al.}

\begin{document}

\title{Dust in the inner regions of debris disks around A stars}


\author{R.L. Akeson\altaffilmark{1}, 
D.R. Ciardi\altaffilmark{1},
R. Millan-Gabet\altaffilmark{1}, 
A. Merand\altaffilmark{2,3}, 
E. Di Folco\altaffilmark{4}, 
J.D. Monnier\altaffilmark{5},
C.A. Beichman\altaffilmark{1},
O. Absil\altaffilmark{6},
J. Aufdenberg\altaffilmark{7},
H. McAlister\altaffilmark{2}, T. ten Brummelaar\altaffilmark{2},
J. Sturmann\altaffilmark{2}, L. Sturmann\altaffilmark{2},
N. Turner\altaffilmark{2}}


\altaffiltext{1}{Michelson Science Center, Caltech, Pasadena, CA 91125}
\altaffiltext{2}{Center for High Angular Resolution Astronomy, Georgia State University, Atlanta, GA 30302}
\altaffiltext{3}{current address: European Southern Observatory, Alonso de Cordova 3107, Casilla 19001, Vitacura, Santiago 19, Chile}
\altaffiltext{4}{Observatoire de Geneve, Universite de Geneve, Chemin 
des Maillettes 51, 1290 Sauverny, Switzerland}
\altaffiltext{5}{Department of Astronomy, University of Michigan, Ann Arbor, MI48109}
\altaffiltext{6}{LAOG, CNRS and Universite Joseph Fourier, BP 53, F-38041, Grenoble, France}
\altaffiltext{7}{Physical Sciences Department, Embry-Riddle Aeronautical University, Daytona Beach, FL 32114}

\begin{abstract}

We present infrared interferometric observations of the inner regions
of two A-star debris disks, \bleo\ and \zlep, using the FLUOR
instrument at the CHARA interferometer on both short (30 m) and long ($>$200 m)
baselines. For the target stars, the short baseline visibilities are
lower than expected for the stellar photosphere alone, while those of
a check star, \dleo, are not.  We interpret this visibility offset of
a few percent as a near-infrared excess arising from dust grains
which, due to the instrumental field of view, must be located within
several AU of the central star.  For \bleo, the near-infrared excess
producing grains are spatially distinct from the dust which produces
the previously known mid-infrared excess.  For \zlep, the
near-infrared excess may be spatially associated with the mid-infrared
excess producing material.  We present simple geometric models which
are consistent with the near and mid-infrared excess and show that for
both objects, the near-infrared producing material is most consistent
with a thin ring of dust near the sublimation radius with typical
grain sizes smaller than the nominal radiation pressure blowout
radius.  Finally, we discuss possible origins of the near-infrared
emitting dust in the context of debris disk evolution models.

\end{abstract}


\section{Introduction}

The list of main sequence stars known to have circumstellar material
in the form of debris disks has been greatly expanded over the last few
years by surveys at longer wavelengths and most recently from
Spitzer observations \citep[see e.g., the review by][]{mey07}.
Given the size and distribution of dust in these disks, the grains are
expected to have short lifetimes. Therefore, it is
generally believed that the dust is not remnant from the star
formation process, but is generated through collisions of
larger bodies.  
The majority of known debris disks have cold ($<$100 K) material
located tens of AU from the central star in an analog of our
own Kuiper belt.  In some cases, this material extends to
1000 AU.  A small fraction \citep{rie05,bei06} have warmer dust
located within 10 AU of the central star.  

The distribution of material in a debris disk is a balance of
collisions, radiation pressure, Poynting-Robertson (PR) drag and the
dynamical influence of any large bodies in the system.  In order to
constrain models of these systems, the dust spatial extent and grain size
distribution must be measured.  Observations of optical and
near-infrared scattered light have provided the most detailed overall
picture of the dust distribution.    However, these
scattered light observations do not have sufficient resolution to
characterize the material closest to the star, and this is where
infrared interferometry can provide a unique constraint.

Although many of these sources do not show a clear near-infrared
excess in their spectral energy distribution (SED), limits set by spatially
unresolved broadband photometry are generally not better than a few to
several percent.  A small, warm dust
component could be present if dust generated by collisions migrated
close to the star or was produced by bodies in close orbits.  If
located within a few AU of the central star, this dust would be at
temperatures which would produce near-infrared emission and small
grains would produce scattered light.  Detection of (or stringent
limits on) warm dust will characterize the inner portions of these
debris disks.  The spatial resolution of infrared interferometry can
be exploited to probe for warm dust in these systems.  On long
baselines ($>$ 100 meters) the central star is resolved and the
visibility is primarily a measure of the stellar photospheric size.
On shorter baselines ($<$ 50 meters) the photosphere is mostly
unresolved and if the measured visibilities have high accuracy, 
other emission components can be detected
by looking for deviations from the visibility
expected for the stellar photosphere.  Any resolved or incoherent
emission will {\em decrease} the measured visibility from the stellar
value.

Teams using the Palomar Testbed Interferometer \citep{cia01} and the Center
for High Angular Resolution Array \citep{abs06,dif07a,abs08} have detected
near-infrared extended emission around known debris disk systems,
including Vega, the prototype debris disk.  While a near-infrared excess
was not known through broadband spectral modeling, the
interferometrically detected near-infrared excess was
consistent with the photometric uncertainties.  Observations
of other debris disk sources revealed a small near-infrared excess
flux around $\tau$ Ceti and $\zeta$ Aql \citep{dif07a,abs08}.
In all these systems, the near-infrared excess is consistent
with emission from an inner, hot dust component, although
for $\zeta$ Aql, a binary companion is also a likely origin.

In this paper we present infrared interferometry observations of
two known debris disk systems, the A-type stars, \bleo\ and \zlep.
The interferometry observations, including
determination of the stellar angular diameter, and 
mid-infrared imaging for \bleo\
are presented in \S \ref{obs}.  Possible origins for the observed
visibility deficit are discussed in \S \ref{vis}.  In \S \ref{discuss},
we discuss the distribution of the excess producing grains and
in \S 5 the origin of these grains.  Our conclusions are
given in \S 6.

\section{Observations and data analysis}
\label{obs}

\subsection{Targets}

The targets were chosen from the sample of known debris disk systems
with the V and K brightness as the primary selection criteria.
Table \ref{tab:prop} lists the target and check star stellar properties.

\begin{table*}[ht!]
\begin{center}
\begin{tabular}{lrrr} \tableline
Parameter & $\beta$ Leo & $\delta$ Leo & $\zeta$ Lep \\ \tableline
HD number & 102647 & 97603 & 38678 \\
Spectral type\tablenotemark{1} & A3Va & A5IVn & A2Vann \\
Distance\tablenotemark{2} (pc) & 11.1 $\pm$0.11 & 17.7$\pm$ 0.26 & 21.5 $\pm$0.32 \\
Radius\tablenotemark{3} (R$_{\odot}$) & 1.58 $\pm$ 0.018 & 2.17 $\pm$ 0.073 & 1.60 $\pm$ 0.11 \\
T$_{\rm eff}$ & 9020\tablenotemark{4} & 8296\tablenotemark{1} & 9910\tablenotemark{1}\\
Luminosity\tablenotemark{5} (L$_{\odot}$) & 11.5 $\pm$ 1.1 & 15.5 $\pm$ 1.8 & 17.0 $\pm$ 2.3 \\
v sin i (km/sec) & 110\tablenotemark{4} & 173\tablenotemark{6} & 245\tablenotemark{4} \\ 
\tableline
\end{tabular} 
\tablenotetext{1}{NASA Stars and Exoplanet database: http://nsted.ipac.caltech.edu}
\tablenotetext{2}{Distances taken from Hipparcos \citep{per97}; we note a more
recent reduction of Hipparocs data \citep{van07} has yielded new distances which
are within 1 $\sigma$ of those listed here.  We use the older values for
consistency with previous work.} 
\tablenotetext{3}{this work}
\tablenotetext{4}{\citet{che06}}
\tablenotetext{5}{calculated from the radius and effective temperature}
\tablenotetext{6}{\citet{rie05}}
\caption{Stellar properties of the sources
\label{tab:prop}}
\end{center}
\end{table*}

\bleo\ was identified as having an infrared excess from IRAS
observations \citep{aum91}.  Mid-infrared imaging has not resolved the
disk \citep[][\S \ref{michelle}]{jay01} although differences between the IRAS and ISO
fluxes led \citet{lau02} to suggest that the disk emission may be
somewhat extended in the ISO beam (52\arcsec\ aperture).
\citet{che06} obtained Spitzer IRS spectra of \bleo\ and found a
featureless continuum spectra consistent with dust at $\sim$120~K located 19
AU from the central star.

\zlep\ was also identified as a debris disk by \citet{aum91} and
has an unusually high dust temperature ($>$300~K) \citep{aum91,che01}.
Recent work by \citet{moe07} resolved the excess at 18~$\mu$m and
their model comprises two dust bands extending from 2 to 8 AU.
As with \bleo, the Spitzer IRS spectrum for \zlep\ is featureless \citep{che06}.

\subsection{CHARA Observations}

Observations were conducted with the FLUOR fiber-optics beam
combiner at the Center for High Angular Resolution Array (CHARA)
operated by Georgia State University.  CHARA is a long-baseline,
six-element interferometer with direct detection instruments that work
at optical to near-infrared wavelengths \citep{ten05}.  These FLUOR
observations were taken in the K' band and have an effective central
wavelength of 2.14 microns.  The FLUOR beam combiner produces high
precision visibilities by interfering the inputs from two telescopes
after spatial filtering through single-mode fibers \citep{cou03}.  In
this design, wavefront aberrations are converted to photometric fluctuations
which are corrected by simultaneous measurement of the fringe and
photometric signals from each telescope.

\bleo\ and a check star \dleo\ were observed on 3 nights in 2006 May 
and \zlep\ on 2 nights in 2006 October and November.  A check star for
\zlep\ was observed but due to its lower K band flux, these data were
not useful and are not included here.  Observations of the targets and
check star were interleaved with calibration observations to determine
the instrument response function, also called the system visibility.  The
check star is an additional target with roughly the same
properties as the main target, but no known excess emission at any
wavelength.  Observations of the check star are processed in the same
way and with the same calibrators as the main target and serve as a measure of
systematic effects in the data.  The calibrators used, along with
their adopted diameters are given in Table \ref{tab:calib}.

\begin{table*}[ht!]
\begin{center}
\begin{tabular}{llll} \tableline
Calibrator & Diameter (mas) & Target & Diameter reference\\ \tableline
70 Leo & 0.770 $\pm$ 0.015 & \bleo, \dleo & SB relation, \citet{ker04b} \\
$\zeta$ Vir & 0.760 $\pm$ 0.015 &  \bleo, \dleo & SB relation, \citet{ker04b} \\
IRC 10069 & 1.342 $\pm$ 0.07 & \zlep & \citet{mer05} \\
$\eta$ Lep & 0.940 $\pm$ 0.020 & \zlep & SB relation, \citet{ker04b} \\
HR 1965 & 1.272 $\pm$ 0.017 & \zlep & \citet{mer05} \\
HR 1232 & 0.920 $\pm$ 0.020 & \zlep & SB relation, \citet{ker04b} \\ \tableline
\end{tabular}
\caption{The calibrators used for the CHARA observations.  The calibrator
sizes are derived using optical and infrared photometry and the
surface brightness (SB) relation from \citet{ker04b} or taken
from \citet{mer05}.
\label{tab:calib}}
\end{center}
\end{table*}

The FLUOR data consist of temporally modulated fringes over an optical
path difference (OPD) of 170 microns, centered around the zero
OPD. The coherence length (fringe packet size) in the K' band is of
order 11 fringes, or approximately 25 microns. In addition to
the fringe signal, FLUOR records simultaneous photometric channels, in
order to allow the correction of scintillation noise and coupling
variations in the input single mode fibers. The photometric correction
and flux normalization were done using the numerical methods described
in \citet{cou97}. Once the fringe signal was
recovered, we estimated the squared visibilities of individual frames
as the integrated power in the frequency domain.

To estimate the fringe power, we used a time/frequency transform, a
Morlet wavelets transform, instead of the classical Fourier approach
\citep{cou97}.  The classical Fourier method extrapolates the power
under the fringe peak using data collected at frequencies outside the
fringe peak \citep{mer06}.  This approach works well if the readout noise is white.
The wavelets approach isolates the fringe signal in the OPD and in the
frequency domains \citep[as described in][]{ker04}, allowing a measure
of the off-fringe power at all frequencies and therefore a direct measurement of
the background noise for each scan.  The isolation of the fringe
signal in OPD is possible because the modulation length used (170
microns) is much larger than the coherence length (approximately 25
microns) and the background noise is measured using the portion of the
scan situated more than 50 microns on each side of the fringe packet
(i.e. four times the coherence length). 

The background noise arises from 3 components: the
photometric variation residuals (after photometric correction), the
photon shot noise and the detector readout noise. The first component
is only present at very low frequencies, since fringes are acquired at a
frequency (100Hz) higher than the scintillation and coupling variations
(typically 25Hz at CHARA) and because the photometric correction is
very efficient. The second component (photon shot noise) is white
noise. The third component, readout noise, is less predictable and can
have transients or peaks at discrete frequencies (electronic noise).
As the wavelet approach directly measures the background component
from the data, there are fewer residuals than in the Fourier method
where the noise estimate is approximate.  For the FLUOR data, the
wavelet method improved the consistency of the results, although
the basic results are the same between the two methods.

Finally, the final squared visibility estimate and the one sigma uncertainty
for a given batch of frames are obtained by the average and
standard deviation of the bootstrapped average, as described in
\citet{ker04}. The calibrated target data obtained using this reduction method
are given in Table \ref{tab:obs}.

\begin{table*}[ht!]
\begin{center}
\begin{tabular}{lrrrrr}\tableline
Object & MJD & Baseline(m) & Pos Angle (deg) & V$^2$ & $\sigma$ \\ \tableline
$\beta$ Leo &  53856.226 &  32.531 & -12.980 &  0.9487 &  0.0219 \\ 
&  53856.270 &  33.234 & -21.034 &  0.9001 &  0.0269 \\ 
&  53856.309 &  33.801 & -26.735 &  0.9285 &  0.0204 \\ 
&  53864.185 & 313.083 &  74.485 &  0.0679 &  0.0079 \\ 
&  53865.185 & 312.858 &  74.288 &  0.0503 &  0.0040 \\ 
&  53865.236 & 293.126 &  68.316 &  0.0897 &  0.0032 \\ \tableline
$\delta$ Leo &   53856.248 &  33.823 & -20.956 &  0.9726 &  0.0393 \\ 
&  53856.290 &  34.069 & -27.244 &  1.0025 &  0.0173 \\ 
&  53856.328 &  33.940 & -31.660 &  1.0356 &  0.0482 \\ 
&  53864.233 & 286.055 &  62.950 &  0.2206 &  0.0253 \\ 
&  53865.215 & 295.815 &  65.870 &  0.1790 &  0.0088 \\ \tableline
$\zeta$ Lep &  54040.479 & 218.336 & -57.363 &  0.7543 &  0.0713 \\ 
&  54040.487 & 223.077 & -57.415 &  0.9403 &  0.1002 \\ 
&  54040.506 & 232.879 & -57.237 &  0.5783 &  0.0772 \\ 
&  54045.475 &  24.739 & -22.258 &  0.9742 &  0.0209 \\ 
&  54045.498 &  26.062 & -26.679 &  0.9524 &  0.0235 \\ 
&  54045.518 &  27.252 & -29.651 &  0.9696 &  0.0272 \\ \tableline
\end{tabular}
\caption{The calibrated visibility observations from CHARA.
\label{tab:obs}}
\end{center}
\end{table*}

\subsection{Stellar size and visibility deficit}
\label{visdef}

\begin{figure*}[ht!]
\includegraphics[angle=270,scale=0.5]{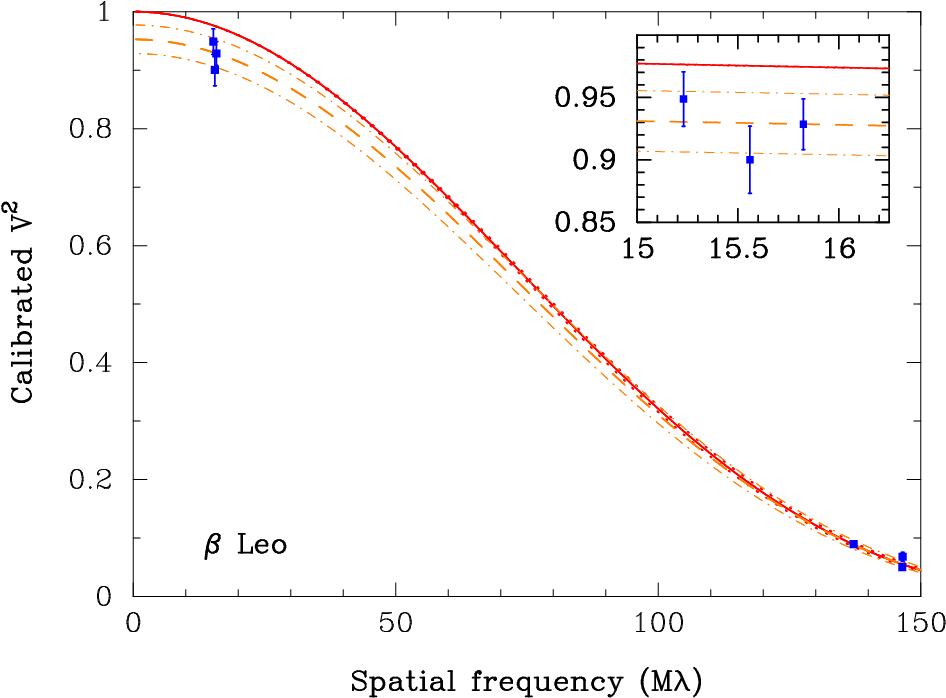}
\includegraphics[angle=270,scale=0.5]{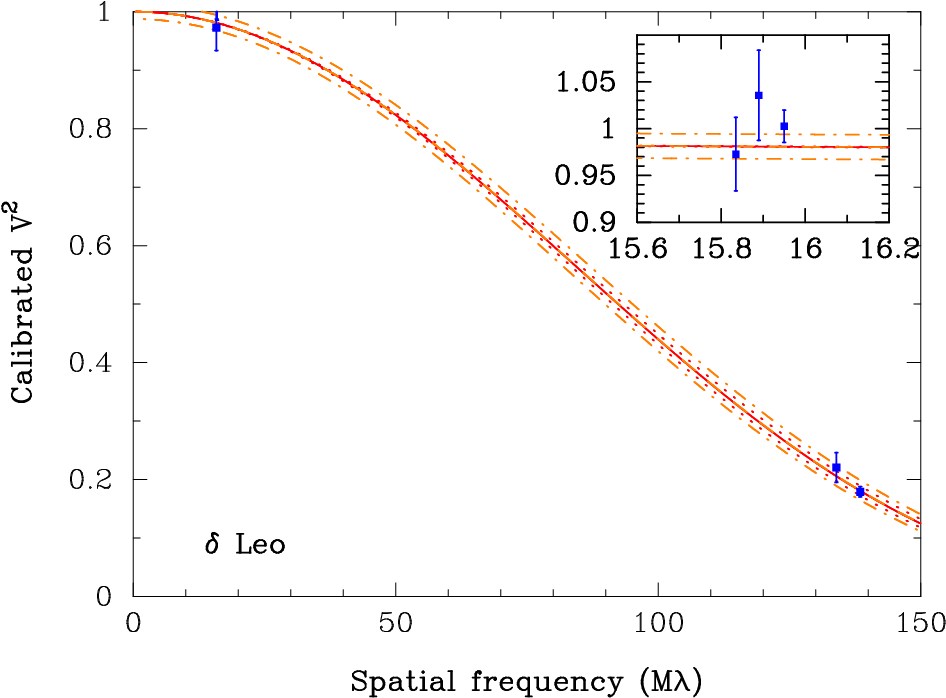}
\includegraphics[angle=270,scale=0.5]{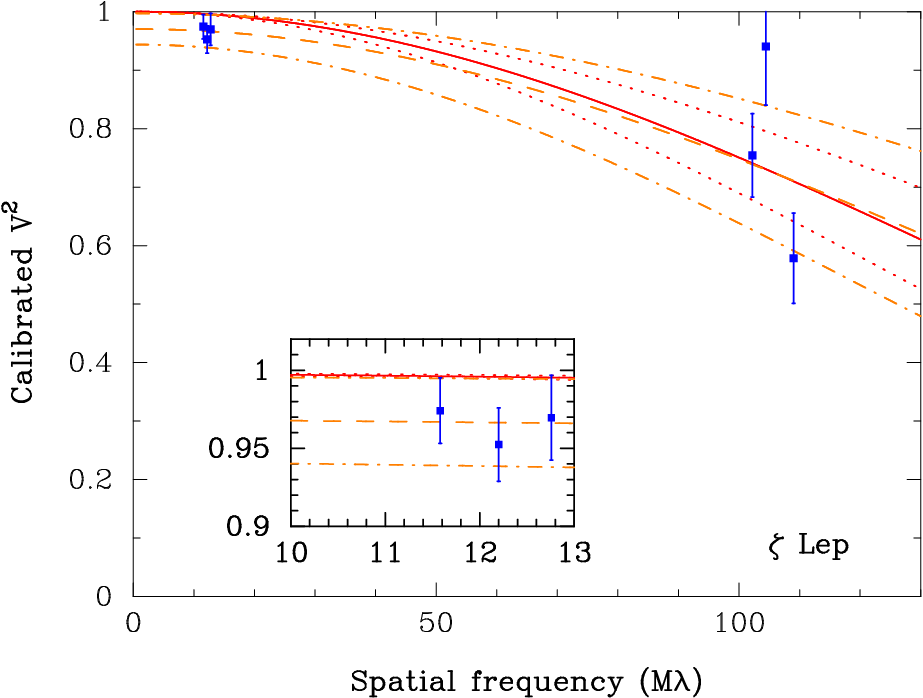}
\caption{The measured visibilities and errors (points) for
\bleo\ (top, left), \dleo\ (top, right) and \zlep (bottom, left).
For each object, the
visibility curves for a uniform disk fit only to the long
baseline data (solid line with dotted line errors) and
for a uniform disk plus an incoherent flux component fit to all the data
(dashed
line with dot-dash line errors) are shown.
\label{fig:vis}}
\end{figure*}

If the measured visibilities were due entirely to a resolved stellar
disk, both the short and long baseline data would be well-fit with a
single uniform disk.  However, as shown in Figure \ref{fig:vis}, the
visibility measured on the short baseline for \bleo\ and \zlep\ is
lower than expected from the stellar size fit on the long baseline.
Fitting a single stellar size to both baselines yields a very poor fit
as measured by $\chi^2_r$ in comparison to the single-baseline only fits
for the target stars, while the single-component fit to both baselines for 
the check star, \dleo,
is good (Table \ref{tab:UDfit}).  Any additional flux component
within the field of view will decrease the measured visibility and
will therefore make the model more 
consistent with the short-baseline data.  A partially
resolved emission component will increase the discrepancy between
the long and short baseline visibilities as it would be more resolved,
and therefore have lower visibility, on the long baselines.  An over
resolved, i.e. incoherent, source of emission will produce the same
fractional decrease in visibility for all baselines.  For the simple
case of a star and an incoherent component, the measured visibility, $V_{\rm meas}$, is
\begin{equation} 
V_{\rm meas}^2 = \left(\frac{V_{\rm star}*f_{\rm star}}{f_{\rm star}
+ f_{\rm incoh}}\right)^2 
\end{equation} 
where $V_{\rm star}$ is the visibility of the stellar photosphere and
$f_{\rm star}$ and $f_{\rm incoh}$ are the fractional stellar and incoherent
component fluxes.  The visibility used here is a normalized quantity
such that an unresolved source has $V=1$ while an incoherent
(i.e. completely resolved) source has $V=0$.  We fit a single uniform
diameter plus an incoherent emission contribution to both baselines,
which gives a lower $\chi^2_r$ for \bleo\ and \zlep\ than the uniform
disk by itself.  The visibility deficit for \zlep\ is a tentative
detection as the V$^2$ predicted from the stellar size, 0.996$\pm$0.001,
is only 2.3 $\sigma$ from the average measured visibility,
0.966 $\pm$ 0.013 and the stellar size uncertainty is much larger as the 
star is smaller and fainter than the other targets.  We note that the
stellar size for \zlep\ from data reduced using the classical Fourier
approach is the same as for the wavelets approach, despite the 
scatter in the long baseline data.
Further observations are needed for confirmation of the visibility
deficit of \zlep.  The best-fit incoherent component corresponds to an excess flux of
2.7 $\pm$ 1.4 Jy for \bleo\ and 0.47 $\pm$ 0.41 Jy for
\zlep.
For \dleo\ the uniform disk fit is adequate,
suggesting no visibility deficit on the check star and no substantial
systematics in the observing or data reduction process.

\begin{table*}[ht!]
\begin{center} 
\footnotesize
\begin{tabular}{lccccccccc}\tableline
Object & \multicolumn{2}{c}{Uniform disk} & \multicolumn{2}{c}{Uniform disk} & \multicolumn{2}{c}{Uniform disk} & \multicolumn{2}{c}{Uniform disk + incoherent flux}\\
& \multicolumn{2}{c}{all data} & \multicolumn{2}{c}{long baselines} & \multicolumn{2}{c}{short baselines} & \multicolumn{2}{c}{all data}\\
& Diam.(mas) & $\chi^2_r$ & Diam.(mas) & $\chi^2_r$ & Diam.(mas) & $\chi^2_r$ & Diam.(mas) & Inc. flux & $\chi^2_r$ \\ \tableline
$\beta$ Leo & 1.332 $\pm$ 0.014 & 3.8  & 1.332 $\pm$ 0.009 & 2.1 & 2.289 $\pm$ 0.31 & 0.9 & 1.323 $\pm$ 0.013 & 0.024$\pm$0.013 & 1.6 \\
$\delta$ Leo & 1.148$\pm$0.025 & 0.8 & 1.149 $\pm$0.012 & 0.5 & 0.0 $\pm$ 1.17 & 0.5  & 1.149$\pm$0.022 & 0.0$\pm$0.006 & 1.0 \\
$\zeta$ Lep & 0.70$\pm$0.15 & 2.5 & 0.69 $\pm$0.09 & 3.7 & 2.0 $\pm$ 0.65 & 0.2 & 0.66 $\pm$0.14 & 0.015$\pm$0.013 & 2.0 \\ \tableline
\end{tabular}
\caption{Uniform diameter and incoherent flux fit to data
\label{tab:UDfit}}
\end{center}
\end{table*}

From the uniform disk fit, we can calculate a limb-darkened angular
and physical diameter for these stars using the
formula from \citet{han74}
\begin{equation}
\frac{\theta_{\rm LD}}{\theta_{\rm UD}} = \left(\frac{1-\mu_{\lambda}/3}
{1-7\mu_{\lambda}/15}\right)^{1/2},
\end{equation} 
where the coefficient $\mu$ depends on the effective temperature and
is taken from \citet{cla95}.  The difference between 
the limb darkened and uniform
disk diameters is less than 2\% for our stars, with ratio values ranging 
from 1.011 to 1.014.  The uniform disk and limb-darkened diameters and the
derived stellar radii are given in Table \ref{tab:size}, where the uniform
diameter is taken from the stellar + incoherent component model.  
These limb-darkened diameters agree with values calculated
from the surface-brightness relation of \citet{ker04}
of 1.35, 1.17 and 0.73 mas for \bleo, \dleo, and \zlep\
respectively.

We note that these limb-darkening parameters are appropriate for
slowly rotating stars, which is violated by the values for $v \sin i$
given in Table \ref{tab:prop}. \citet{auf06} find limb-darkening
corrections 2.5 times higher for Vega, an A0 star rotating at 275
km/sec.  As our data are insufficient to separately derive the
limb-darkening or rotational velocity, we use the low rotation rate
coefficients to allow for comparison to other works, but note that
even at 2.5 times higher, the limb darkening corrections would be 4\%,
still much too small to explain the difference between the short and
long baseline sizes given in Table \ref{tab:UDfit}.

\begin{table*}[ht!]
\begin{center}
\begin{tabular}{lcccc} \tableline
Object & Uniform disk diam. & LD coeff & Limb darkened diam. & Stellar radius \\
& (mas) & & (mas) & (R$_{\odot}$) \\ \tableline
\bleo\ & 1.323 $\pm$ 0.013 & 1.012 & 1.339 $\pm$ 0.013 & 1.54 $\pm$ 0.021 \\
\dleo\ & 1.149 $\pm$ 0.022 & 1.014 & 1.165 $\pm$ 0.022 & 2.14 $\pm$ 0.040 \\
\zlep\ & 0.66 $\pm$ 0.14 & 1.011 & 0.67 $\pm$ 0.14 & 1.50 $\pm$ 0.31 \\
\tableline
\end{tabular}
\caption{Measured uniform disk and limb-darkened diameters
\label{tab:size}
}
\end{center}
\end{table*}

The diameter of \bleo\ has been previously measured with
interferometry observations.  \citet{han74} obtained a limb-darkened
diameter of 1.33 $\pm$ 0.1 mas at a wavelength of 4430 \AA\ with the
Narrabri intensity interferometer,
while \citet{dif04} measured 1.449 $\pm$ 0.027 mas at 2.17 $\mu$m with
the VLTI, which is inconsistent with our diameter at the 3.7$\sigma$ level.  
However, the \citet{dif04} fit did not include an
incoherent component.  If we include the \citet{dif04} data in our
two component fit, both the stellar diameter and incoherent flux level
change by less than 0.2 $\sigma$, thus the VLTI and CHARA data are
consistent.

\subsection{Mid-infrared imaging}
\label{michelle}

Mid-infrared imaging observations of $\beta$~Leo were made on 2006 March
8 (UT) using the Mid-Infrared Echelle Spectrometer \citep[MICHELLE;
][]{glasse93} on the Gemini North 8-meter telescope. MICHELLE utilizes a
$320 \times 240$ pixel Si:As blocked impurity band detector, with a
spatial scale of $0\farcs1$ pixel$^{-1}$.  Imaging was obtained in the
Qa filter ($\lambda_c = 18.1$\micron, $\Delta\lambda = 1.9$\micron) with
a standard off-chip 15\arcsec\ ABBA chop-nod sequence and a chop
position angle of $30\deg$ E of N.  Two image sequences of $\beta$~Leo
were taken with 30 ms frametimes and a total on-source integration time
of 325 s per image.  Prior to and following the $\beta$~Leo
observations, HD109511 (K0, $F_{18\mu\rm{m}} \approx 1.4$ Jy) was
observed with the same observing sequence to serve as a point spread
function and flux density calibrator.  The data were reduced with
custom-written IDL routines for the MICHELLE data format.

\bleo\ appears unresolved in comparison to the calibrator.  At 18.5
\micron, the excess for $\beta$~Leo is $\sim 0.3$ Jy
\citep{che06} and at 19 AU from the star, the radius inferred by
\citet{che06} for the mid-infrared emitting material, we measured
an rms dispersion of the background in the $\beta$~Leo images of $\sim
0.55$ mJy pixel$^{-1}$.  In \S \ref{geom}, we will use this limit to
constrain the radial extent of the mid-infrared emitting material.

\section{Origin of the visibility deficit}
\label{vis}

In this section we discuss the possible origins of the 
visibility deficit.

\subsection{Companion}

A companion anywhere within the 0\farcs8 (FWHM) field-of-view (FOV)
will lower the measured visibility.  A companion within the fringe
envelope (roughly 25 milliarcsec for these observations) will produce
a visibility modulation which is a function of the binary flux ratio
and separation and the projected baseline length and position angle.
A companion outside this separation range but within the field-of-view
will contribute incoherent flux and the visibility decrease will be
the same fraction on all baselines.  The flux ratio of a companion
which would produce the measured visibility is the incoherent fraction
listed in Table \ref{tab:UDfit}, which corresponds to $\Delta$K = 4.0
$\pm$ 0.9 for \bleo\ and $\Delta$K = 4.5 $\pm$ 1.4 for \zlep.  These
flux differences would be produced by a main sequence star of spectral
type M0 for \bleo\ and M2 for \zlep.

Neither star has a known companion within a few arcsec of the primary
star.  The Washington Double Star (WDS) catalog lists 3 companions for
\bleo, located from 40\arcsec\ to 240\arcsec\ from the primary (far
outside the FOV) with V magnitude differences of 6.3 to 13
\citep{wor97}.  None of these stars could affect the interferometry
observations due to the large angular separation.  \zlep\ has no
listed companions in the WDS.  Both objects have been imaged in the
mid-infrared \citep[][\S \ref{michelle}]{jay01,moe07} with no
companion detected.  In our MICHELLE/GEMINI data, the Q-band magnitude
difference for a point source which can be ruled out is 2.5 mag within
0\farcs5 and 4 mag from 0\farcs5 to 0\farcs8. These data are
sufficient to detect a possible companion between 0\farcs5
to 0\farcs8 around \bleo\ for the
derived companion spectral type of M0.

The strongest constraints on close ($<$1\arcsec) companions come from
the {\it Hipparcos} measurements.  \bleo\ was observed 64 times over
3.0 yrs with final positional uncertainties of 0.99 mas (RA) and
0.52 mas (dec), and \zlep\ was observed 117 times over 3.1 yrs with
uncertainties of 0.51 mas (RA) and 0.41 mas (dec) \citep{per97}.  As
neither source was detected to have any astrometric motion by {\it
Hipparcos}, these uncertainties can be used to place limits on any
stellar companions.  Using the secondary stellar types inferred from
the flux ratios, the companion stellar masses would be approximately
0.5 M$_{\odot}$ for \bleo\ and 0.4 M$_{\odot}$ for \zlep.  As the
astrometric signature increases with orbital distance, the astrometric
uncertainty from the {\it Hipparcos} data sets a lower limit to the excluded
periods, while the sampling duration sets the limit for longer-period
companions.  To estimate the shortest period companion which the {\it
Hipparcos} data could detect, we assumed a mass for each primary of 2.0
M$_{\odot}$ and quadratically combined the positional uncertainties to
obtain astrometric uncertainties of 1.12 mas for \bleo\ and 0.65 mas
for \zlep.  Setting a threshold of 5$\sigma$ to account for the
uneven time sampling, the minimum detectable separations are 0.25 AU
(\bleo) and 0.34 AU (\zlep), which correspond to periods of 32 days
and 51 days respectively.  

The detection of longer-period companions is limited by the overall
time span of the {\it Hipparcos} data.  The orbital period and astrometric
signature of a companion located at the edge of the FOV would be 5.5
yrs and 250 mas for \bleo\ and 14.5 yrs and 200 mas for \zlep. For
\bleo, the {\it Hipparcos} data samples half a period and would be
sufficient to detect such a companion. For \zlep, the {\it Hipparcos}
data would sample 20\% of the orbital period. For a circular orbit, the
deviation of this arc from a best-fit straight line would be 12 mas,
detectable with the 0.65 mas uncertainty, but detecting some phases of
an elliptical orbit would be more difficult.  A very long period
companion with the relevant magnitude difference could have escaped
detection if the orbit is inclined on the sky such that companion is
currently too close to the primary (within 0\farcs5) for detection by
imaging.  One probe of such a very long period orbit is the proper motion as
a function of time.  \citet{gon01} combined proper motion data from
ground-based catalogs starting in the 1930's with the {\it Hipparcos}
data.  For \bleo\ and \zlep, the combined proper motions were within
the uncertainties of the {\it Hipparcos} proper motions, and both
stars were classified as having no companions within 10\arcsec.

Any companion closer than the short-period limit derived above would
produce a substantial radial velocity signature.  Using the
inclination angles derived in \S \ref{oblate}, a companion at the
short period limits above would produce a radial velocity of 8
km/sec for \bleo\ and 12 km/sec for \zlep.  \citet{gal05} made
measurements of \bleo\ with an uncertainty of 137 m/s, more than sufficient to
detection such a large signature, however, the time sampling covered
only a few hours and is not sufficient to rule out companion periods
of tens of days.  Observations of \zlep\ \citep[][e.g.]{gre99} have
also been made with sufficient precision, but not sufficient
time sampling to find a companion with a period of a many days.

In summary, neither target star has a known companion within the CHARA
FOV and {\it Hipparcos} measurements rule out companions with
periods from tens of days to several years.  A very close companion
(periods less than tens of days, separations less than 0.35 AU) can
not be ruled out in either case, but would produce an easily
detectable ($>$ 5 km/sec) radial velocity signature.  Although we can
not definitively rule out a companion as the source of the flux
decrement, it is unlikely given the above constraints on period and
magnitude difference. A less massive companion would produce a smaller
flux decrement, which would require another flux component in the
system.  Given the small phase space remaining for an undetected companion
and the fact that the two mid-infrared excess sources (\bleo\ and \zlep) 
have a near-infrared visibility decrement, while \dleo\ with no mid-infrared
excess does not, we proceed with the hypothesis that the flux
decrement does not arise from a companion.

\subsection{Stellar rotational oblateness}
\label{oblate}

Our analysis of the visibility deficit on the short baseline relies on
knowledge of the stellar size from the longer baselines.   If the
star is oblate due to rotation, the predicted size on the
short baseline may be incorrect as the short and long baselines
are nearly orthogonal (Table \ref{tab:obs}).  We 
can calculate the maximum possible effect by assuming the
short stellar axis is aligned with the longer baseline, which would place
the longer axis along
the short baseline, producing lower visibilities.  
We calculate the ratio of stellar radii, $X_R$ from
\begin{equation}
X_R = \frac{R_{\rm pol}}{R_{\rm eq}} = \left(1+\frac{v_{\rm eq}^2 R_{\rm eq}}{2GM}\right)^{-1}
\end{equation}
where $R_{\rm pol}$ and $R_{\rm eq}$ are the polar and equatorial radii, $v_{\rm eq}$
is the equatorial velocity, $G$ is the gravitational constant and $M$ is the
stellar mass \citep{des02}.  For the most conservative calculation,
we take $v_{\rm eq}$ to be the maximum equatorial velocity inferred by
\citet{roy07} of a survey of A stars, which are grouped by sub-class.
These velocities are 300 km sec$^{-1}$ for \bleo\ and \zlep\ and 280 km sec$^{-1}$
for \dleo.  The resulting oblateness is corrected for viewing
angle by deriving $i$ from the measured $v \sin i$ and the assumed
$v_{\rm eq}$ and approximating the stellar shape as an ellipsoid (Table \ref{tab:oblate}).
The observed stellar radii ratio $X_{\rm obs}$ is then given by
\begin{equation}
X_{\rm obs} = \frac{X_R}{(1-(1-X_R^2)\cos^2 i)^{1/2}}
\end{equation}

Starting with the derived stellar size on the long baseline
($\theta_{\rm long}$, see Table \ref{tab:UDfit}), 
we calculated the V$^2$ that would be measured on the
short baseline (V$^2(\theta_{\rm long})$).  
We then applied
the observed oblateness factor, $X_{\rm obs}$ to find
the maximum possible angular diameter, $\theta_{\rm long}/X_{\rm obs}$, 
and recalculated
the V$^2$ for the short baseline (V$^2(\theta_{\rm long}/X_{\rm obs}))$.
Because these angular sizes are at best marginally
resolved on the short baseline, the change in visibility
is less than 1\% in all cases, even if the apparent angular size
changes by 20\%, as predicted for \zlep.  
For comparison, we also list the short baseline size, $\theta_{\rm short}$ 
from Table \ref{tab:UDfit}.
The measured visibility on the short baseline, V$^2_{\rm measured}$
is significantly lower than either V$^2(\theta_{\rm long})$ or 
V$^2(\theta_{\rm long}/X_{\rm obs})$ for
both \bleo\ and \zlep\ but not for the check star
\dleo\ and thus stellar oblateness can not account for
the measured visibility deficit.  We note
that if rotational axis of the star is aligned such that the 
short stellar axis is along the 
short baseline, then the true visibility decrement is actually
slightly larger than measured.  

As these stars are rotating rapidly, they are also subject to
gravity darkening, which produces a decrease in the effective
temperature from the pole to the equator.  Since the limb-darkening
depends on the effective temperature, this effect is also
linked to the apparent oblateness.  However, this effect
is very small compared to the oblateness derived above.
Using the effective temperature difference found
by \citet{auf06} for Vega, an A0 star, of 2250~K, the 
limb-darkenening correction for the pole is 0.3\% larger than
correction at the equator.  This factor goes against the
rotational oblateness which makes the equatorial radius
larger and even with the factor of 2.5 for a fast
rotating star, is insufficient to explain the
ratios between diameters fit to the long and
short baselines of 1.72 $\pm$ 0.23 for \bleo\ and
2.9 $\pm$ 1.0 for \zlep.

\begin{table*}[ht!]
\begin{center}
\begin{tabular}{llll} \tableline
& \bleo & \dleo & \zlep \\ \tableline
$v \sin i$ (km sec$^{-1}$) & 110 & 173 & 245 \\
assumed $v_{\rm eq}$ (km sec$^{-1}$) & 300 & 280 & 300 \\
i (deg) & 21.5 & 38.1 & 54.7 \\
R$_{\rm eq}$ (R$_{\odot}$) & 1.54 & 2.14 & 1.5 \\
X$_R$ & 0.74 & 0.70 & 0.74\\
X$_{\rm obs}$ & 0.95 & 0.84 & 0.80 \\
$\theta_{\rm long}$ (mas) (Table \ref{tab:UDfit}) & 1.332$\pm$0.009 & 1.149$\pm$0.012 & 0.69$\pm$0.09 \\
$\theta_{\rm long}/X_{\rm obs}$ (mas) & 1.401$\pm$0.009 & 1.368$\pm$0.014 & 0.826$\pm$0.11 \\
$\theta_{\rm short}$ & 2.289 $\pm$ 0.31 & 0.0 $\pm$1.17 & 2.0 $\pm$ 0.65 \\
V$^2$ on short baseline: \\
\quad V$^2(\theta_{\rm long})$ & 0.976$\pm$0.0003 & 0.981$\pm$0.0004 & 0.996$\pm$0.001 \\
\quad V$^2(\theta_{\rm long}/X_{\rm obs})$ & 0.973$\pm$0.0003 & 0.973$\pm$0.0006 & 0.994$\pm$0.002 \\ 
\quad V$^2_{\rm measured}$ & 0.938$\pm$0.015 & 1.001 $\pm$ 0.015 & 0.966 $\pm$ 0.013 \\
\tableline
\end{tabular}
\caption{The calculated maximum visibility change due to rotational oblateness.
The uncertainties in $\theta_{\rm long}/X_{\rm obs}$, V$^2(\theta_{\rm long})$,
and V$^2(\theta_{\rm long}/X_{\rm obs})$ include the uncertainty in the
measured value of $\theta_{\rm long}$ but not the uncertainty in X$_{\rm obs}$,
which is unknown.
\label{tab:oblate}}
\end{center}
\end{table*}

\subsection{Emission and scattering from dust}
\label{dust}

Dust grains within the field of view will produce a near-infrared
excess through thermal emission and scattering.  We assume
that there is no gas in these debris disks and therefore
the inner radial limit for the debris disk is the dust
sublimation radius.  For a sublimation temperature of
1600~K and assuming large grains in thermal equilibrium emitting as blackbodies, 
the sublimation radius is 0.12 AU for \bleo\
and 0.14~AU for \zlep.  The 2 $\mu$m emission will be maximized
for dust at the sublimation temperature, so a lower limit
to the excess luminosity can be estimated following \citet{bry06}
\begin{equation}
\frac{L_{\rm dust}}{L_{*}} = \frac{F_{\rm dust}}{F_{*}}\frac{kT_{\rm dust}^4 (e^{h\nu/kT}-1)
}{h\nu T_{*}^3}.
\end{equation}
where $h$ and $k$ are the Planck and Boltzmann constants.
For a temperature of 1600~K, the fractional dust luminosity is
$2.0 \pm 1.1 \times 10^{-3}$ for \bleo\ and
$9.8 \pm 8.5 \times 10^{-4}$   for \zlep.  For comparison, \citet{che06}
calculated mid-infrared dust luminosities of $2.7 \times 10^{-5}$ and $6.7 \times
10^{-5}$ for \bleo\ and \zlep\ respectively.
However, the much larger near-infrared luminosity does not require substantially
more mass than implied by the mid-infrared excess since, as the 
fractional dust luminosity represents
the fraction of the star as seen by the dust, the calculated 
fractional luminosities are highly sensitive to the dust location.
An estimate of the minimum mass of near-infrared emitting grains can
be calculated using the fractional luminosity and
assuming efficiently emitting grains \citep{jur95},
\begin{equation}
M_{\rm dust} \geq \frac{16 \pi}{3} \frac{L_{\rm dust}}{L_{\ast}} \rho a r^2
\end{equation}
where $\rho$ is the density, $a$ is the grain radius and $r$ is
the distance from the star.  A minimum mass can be calculated
by using the $L_{\rm dust}$ values calculated above for
small dust grains located near the sublimation radius.
For a grain radius of $a = 1 \mu$m, $r$ at the dust
sublimation radius and $\rho \sim 2$ gm~cm$^{-3}$, the
minimum mass of the near-infrared emitting material is 
$5 \times 10^{-9}$ M$_{\oplus}$ for \bleo\ and
$2 \times 10^{-9}$ M$_{\oplus}$ for \zlep . \citet{che06}
derived a mass for the small grains in the mid-infrared
producing material of $4.2 \times 10^{-6}$ for \bleo\ and 
$5.6 \times 10^{-6}$ for \zlep.  So although the near-infrared
excess represents a higher fractional dust luminosity,
this can be produced by a much smaller mass than the mid-infrared
ring.

A ring of hot dust near the sublimation radius is not incompatible
with the incoherent flux model fit in \S \ref{visdef}, as the sublimation
radius is large enough to be resolved on even the short baseline.
Given the relative uncertainty in the
incoherent flux component fit, a component with V$^2 < 0.2$ would fit
within the uncertainty.  
For \bleo, the sublimation radius corresponds to 11
mas and a ring of any width at this radius  has a V$^2 < 0.2$ on all
baselines in our observations. For \zlep, the sublimation radius is at 7 mas
and any ring wider than 1 mas (0.02 AU) produces V$^2 < 0.2$ on 
all baselines.  If the inclination angles are close to the values
inferred in Table \ref{oblate}, these approximations are sufficient. 
Thus, thermal emission from hot dust near the sublimation radius could
produce the measured visibility deficit.

At larger angular scales  than have
been investigated with the interferometer $(\gtrsim 1\arcsec - 10\arcsec)$, 
debris disks are often
detected in scattered light at optical and near-infrared wavelengths
\citep[e.g., AU Mic and Fomalhaut;][]{kal04, kal05} and scattering
from within the field of view of the interferometer ($\lesssim$
0\farcs8) could also produce the observed visibility deficit.
Scattering in the near-infrared will dominate emission for grains at
several hundred degrees, depending on the grain size and composition.
To investigate the scattering from warm dust, we used the debris disk
simulator\footnote{http://aida28.mpia-hd.mpg.de/$\sim$swolf/dds/} 
described by \citet{wol06} which calculates the thermal
emission and scattering given the dust size, composition and
distribution.  For example, small grains uniformly distributed from
1.0 to 4.6 AU (the \bleo\ FOV radius) will produce the observed near-infrared excess given
a total mass of small grains of $1-7 \times 10^{-5}$~M$_{\oplus}$,
depending on the exact size and composition.  This is 
more than 1000 times larger than the minimum mass 
needed to produce the excess from
hot grain emission.

As there is no known evidence for a companion, we contend that thermal
emission and scattering from dust grains is the most likely origin of
the near-infrared excess.  This is also consistent
with our finding that the two sources with a measured visibility deficit have
mid-infrared excess emission while the control star, which has
no known excess does not have a visibility deficit.
In the next section, we explore the
constraints on these grains and discuss possible mechanisms for their
origin.

\section{Dust distribution and small grain origin}
\label{discuss}

\subsection{Dust grain sizes}

Both \bleo\ and \zlep\ have a substantial mid-infrared excess which
has a characteristic temperature much lower than dust which would
produce a near-infrared thermal excess and is therefore further from
the central star.  Many authors \citep[see e.g.][and references therein]
{dom03,wya05} have studied the dynamics of
debris disks similar to our targets and have found that collisions are
dominant over PR drag, i.e grains collide and become smaller before PR
drag significantly decreases the size of their orbits.  Radiation
pressure also plays a role as small grains are subject to removal from
the system.  However, clearing of small grains may not be absolute.
\citet{kri00} modeled the $\beta$ Pic disk, which has a similar
spectral type (A6V) and optical depth to the systems discussed here and
found that although grains at and below the canonical blowout
radius are depleted compared to a purely collisional system, a
population of small grains persists in their model.  For our target stars,
the radiation pressure size limit is $\sim$2~$\mu$m, the PR drag
timescale at 1 AU is 1000 yrs for 10 $\mu$m radius grains and the
collisional timescale for these same grains is 80 years \citep[following
the formula of][]{bac93}.

A second constraint on the dust size is the lack of a significant
silicate feature in the IRS spectrum for either source 
\citep[][see Figure 2]{che06},
although the excess for \bleo\ is not strong enough at 10$\mu$m to
provide as strong a constraint as for \zlep, which has excess emission
at shorter wavelengths.  The lack of a silicate emission feature
requires the grain population to have radii larger than a few microns if
composed of silicates or to be primarily non-silicate.  

\subsection{Modeling approach}
\label{modeling}

As the interferometer data provide only an upper limit to the
visibility and therefore a lower limit to the size of the
near-infrared flux region, the strongest spatial constraint from the
interferometry data is that the dust must be within the FOV.  However,
there is another strong constraint from the measured mid-infrared
excess of these sources.  The dust producing the near-infrared excess
will also produce mid-infrared excess, with the exact flux depending of
course on the dust temperature and opacity.

We now begin to explore various specific models for the
distribution of dust in these systems, and examine whether these
models fit within the constraints provided by the near and
mid-infrared data.    In all models, we assume optically thin
emission for the near and mid-infrared emission. In this
section, we consider the relative contributions of scattering
and emission to the near and mid-infrared excess flux.
For the scattering, we have used the 
debris disk models
of \citet{wol06} to calculate the emission and scattered light flux for
various grain radius and radial distributions and two
example grain compositions.  We have chosen grain compositions which
will produce the featureless mid-infrared spectrum seen in the IRS
data and have substantially different emissivity ratios between the
near and mid-infrared.  These two populations are silicate grains
with radii between 3 and 10 $\mu$m and graphite grains with radii
from 0.1 to 100 $\mu$m.  In both cases, we use a distribution
of grain radii, $n(a) \propto a^{-3.5}$ appropriate for collisionally
dominated disks.
For these toy models, we
concentrated on illustrative cases of dust
radial distributions and did not modify the grain radius distribution
for the effects of radiation pressure.  The possible
presence of small grains is discussed in more detail in \S \ref{geom}.
For each case, the disk mass
was determined by scaling to match the observed near-infrared excess.
These masses are significantly higher than the minimum mass 
derived in \S \ref{dust} as that estimate assumes the flux comes only from
small, hot grains which produce much more near-infrared emission
for the same mass than a distribution of grain sizes and
temperatures can.
In Table \ref{tab:flux} we present the results for the two 
grain populations over 
several radial distributions, listing the 
ratio of emission to scattering at 2 $\mu$m, the excess flux at 10 and 24
$\mu$m and the mass in small grains.  All models have radial
density profiles of $n(r) \propto r^{-1.5}$.

\begin{table*}[ht!]
\begin{center}
\begin{tabular}{|ll|llll|llll|}\tableline
& & \multicolumn{4}{c|}{\bleo} & \multicolumn{4}{c|}{\zlep}\\ \tableline
Radial & Grain & $F_{\rm em}/F_{\rm sc}$ & $F_{10\mu m}$ & $F_{24\mu m}$ & M$_{\rm small gr}$ & $F_{\rm em}/F_{\rm sc}$ & $F_{10\mu m}$& $F_{24\mu m}$ & M$_{\rm small gr}$ \\ 
distribution & type & & (Jy) & (Jy) & M$_{\oplus}$ & & (Jy) & (Jy) & M$_{\oplus}$ \\ \tableline
R$_{\rm sub}$-FOV & silicate & 6.1 & 1.7 & 1.2 & $1\times 10^{-5}$ & 8.5 & 18 & 16 &  $1\times 10^{-4}$ \\
& graphite  & 8.6 & 5.4 & 2.8 & $2\times 10^{-6}$ & 10 & 6.7 & 4.6 & $2\times 10^{-5}$ \\
1.0 AU-FOV & silicate & 0.019 & 231 & 254 & $4\times 10^{-4}$ & 0.02 & 261 & 400 & $4 \times 10^{-3}$ \\
& graphite  & 4.9 & 2.0 & 1.8 & $2 \times 10^{-5}$ & 4.1 & 32 & 35 & $2 \times 10^{-4}$ \\
R$_{\rm sub}$-1.0 AU & silicate & 8.6 & 10 & 3.9 & $1\times 10^{-6}$ & 8.8 & 10.5 & 4.0 & $5 \times 10^{-6}$ \\
& graphite & 9.2 & 4.7 & 2.0 & $1\times 10^{-6}$ & 12 & 4.7 & 2.1 & $ 5\times 10^{-6}$ \\ \tableline
\end{tabular}
\caption{The ratio of emission to scattering flux at 2 $\mu$m, the mass in
small grains necessary to reproduce the observed near-infrared excess and
the 10 and 24 $\mu$m flux for several disk models. The radius of the
FOV  
corresponds to 4.6 AU for \bleo\ and 8.6 AU for \zlep.
\label{tab:flux}
}
\end{center}
\end{table*}

As expected, emission dominates for grains close to the central star
($<$1~AU), while scattering dominates for grains farther away.  The
mass in small dust grains necessary to produce the near-infrared
excess flux is higher for scattering-dominated disks than for 
emission-dominated
disks.  The scattering-dominated cases produce too much mid-infrared
flux, in some cases by more than an order of magnitude.  The 24 $\mu$m
excesses measured by \citet{su06} are 0.46$\pm$0.01 Jy for \bleo\ and
0.53$\pm$0.02 Jy for \zlep\ and the 10 $\mu$m excess from the IRS
spectra are 0.002$\pm$0.004 Jy for \bleo\ and 0.18$\pm$0.01 Jy for
\zlep\ from \citet{che06}.  The emission-dominated disks also produce
too much mid-infrared flux, but not by as large a factor.  As the
models which have substantial near-infrared emission have mid-infrared
fluxes close to the observed values, we assume that emission is the
primary mechanism for producing near-infrared flux.  In the following
sections, we will explore other density distributions and models to
fit both the near-infrared and mid-infrared excesses in detail.

\subsection{Dust grain distributions}

For both stars, we first considered the hypothesis that the grains producing
the near-infrared excess were generated by collisions between larger
bodies in the belt which produces the mid-infrared excess.  These
grains can then be dragged towards the central star via PR drag and
become sufficiently heated to emit at near-infrared wavelengths.  
For a specific theoretical description 
of a disk in which grains
created in collisions in the planetesimal belt migrate inward, we
used the model of \citet{wya05}, who calculated the steady-state
optical depth as a
function of radius. In this model, the disks are
collisionally dominated, but a small fraction of the dust 
created by collisions
in the planetesimal belt migrates inwards due to PR drag and
is subject to collisions as it migrates.
Assuming a single grain size, \citet{wya05} found the optical
depth as a function of radius to be 
\begin{equation} 
\tau_{\rm eff}(r) =
\frac{\tau_{\rm eff}(r_0)}{1 + 4 \eta_0(1-\sqrt{r/r_0})} 
\label{eq:tau}
\end{equation} 
where $\tau_{\rm eff}(r_0)$ is the optical depth of the
planetesimal belt at $r_0$, the radius of the planetesimal belt and
$\eta_0$ is a parameter balancing collisions and PR drag, which had a
value of 2.4 for \bleo\ and 6.7 for \zlep.  For $\eta_0 = 1$ the
collisional lifetime equals the time it takes a grain to migrate to
the star.  We assume optically thin, blackbody grains
distributed with the optical depth given by eq. \ref{eq:tau} and
starting at the sublimation radius.  The value of $\tau_{\rm eff}(r_o)$ is
iterated until the optical depth within the CHARA FOV produces the
observed near-infrared excess.  We then calculate how much
mid-infrared flux would be produced and compare to the measured mid-infrared excess.

For \bleo, the mid-infrared excess spectra is well fit by a grain
temperature T $\approx$120~K, which implies a distance from the star of 19 AU
\citep{che06}, well outside
the FOV of our observations (4.6 AU).  Applying the model in
eq. \ref{eq:tau} with $r_0 = 19$~AU and assuming an
emissivity wavelength dependence of $\lambda^{-2}$ (\S \ref{geom}) produces a 2/10 $\mu$m flux
ratio of 1.6, while the observed ratio, using our detection
and the IRS data of \citet{che06} is $> 140$.
A shallower grain emissivity function with wavelength will produce an even
larger discrepancy between this model and the data.  
Thus, the grains which produce
the near-infrared excess can not come from a smoothly distributed
population generated by collisions in the mid-infrared belt.

\citet{moe07} resolved the 18~$\mu$m emission from \zlep\ and modeled
the distribution as arising from two rings with stellar distances 
from 2-4 and 4-8
AU, which is contained within the CHARA FOV of 8.6 AU.
Using a radius of 4~AU in eq. \ref{eq:tau} produces a 
2/6 $\mu$m excess of 0.8, while the observed excess from
our data and \citet{che06} is 18.  Thus this model is
not a good fit for \zlep\ either.
We note that \citet{moe07} did not resolve
the excess emission from \zlep\ at 10 $\mu$m and concluded that the
dust producing this excess is interior to the resolved 18 $\mu$m rings.
They surmise that the 10~$\mu$m emitting dust is migrating inward by
PR drag from the belts resolved at 18~$\mu$m.

\subsection{Geometric models}
\label{geom}

To further examine the constraints which the near-infrared and
mid-infrared excess place on the dust distribution, we use a geometric
model of the dust distribution.  The data are compared to the models
in a Bayesian approach designed to constrain the range
of valid model parameters, rather than finding a single best-fit model.
The input 
data are: 1) the near-infrared excess within the interferometer FOV
and the visibility limits for this excess, 2) the IRS data from
\citet{che06} and 3) the spectral energy distribution from 2 to 100
$\mu$m from the literature, including a 70 $\mu$m Spitzer-MIPS
measurement (K. Stapelfeldt, private communication).  We have
constructed an SED for each star using photometry from SIMBAD in order
to determine the excess flux.
The stellar template was determined by fitting the optical and
near-infrared photometry to a grid of Kurucz-Lejeune models
\citep{lej97} covering the range of effective temperature and surface 
gravity values
appropriate for the main sequence stellar types of the target
stars.  Both stars are nearby and have photospheric colors consistent
with $A_V = 0$ \citep{che06}.

The basic disk model is an optically thin ring of dust. 
To simplify the calculations, we consider the dust to be geometrically
thin; however, we note that to intercept $\sim$1\% of the starlight,
the dust will need to have a finite vertical height.  At a radial
distance of 0.1 AU, this corresponds to a height $h$ with
$h/r = 0.02$.  Such a vertical height is smaller
than a flared primordial disk at this radius \citep[$h/r=0.09$]{chi97}
and smaller than the value $h/r \sim 0.05$ derived for the $\beta$ Pic
dust disk at larger (r $>$ 15 AU) radii and thus is plausible. 

The excess flux ratio between 2.2 $\mu$m and the shortest IRS
wavelength can be used to set a limit on the wavelength dependence of
the grain emissivity.  For \bleo, F(2.2 $\mu$m)/F(10 $\mu$m) $>140$, 
implying an emissivity decreasing at least as fast as $\lambda^{-1.3}$ if the
emission is from a hot blackbody.  For \zlep, F(2.2 $\mu$m)/F(6
$\mu$m) $\sim$ 18, implying emissivity proportional to at least
$\lambda^{-0.9}$ for hot dust.  
As we wish to examine the range of disk physical
parameters, including the grain size and composition, 
which can reproduce the near and mid-infrared excess
emission, we adopt a analytic approximation for the grain properties
to keep the calculation manageable yet self-consistent. 
We have therefore chosen a power-law representation of the radiative
efficiency $\epsilon$, following work on $\beta$ Pic by \citet{bac92}
and \citet{bac93}. For a given grain radius $a$, the absorption and
emission efficiency is
roughly constant, $\epsilon \simeq 1 - \rm{albedo}$, for radiation 
at wavelengths shorter than a critical
wavelength $\lambda_o$ and decreases for wavelengths longer than
$\lambda_o$ \citep{bac93}.  The relation between the grain radius $a$ and
the critical wavelength depends on the grain composition and shape and
varies from $\lambda_o/a \sim 2 \pi$ for strongly absorbing grains to
$\lambda_o/a \sim 1/2 \pi$ for weakly absorbing grains \citep{bac93}.

We assume that the disk is optically thin to its own radiation,
therefore the stellar radiation is the only input.  Radiation from
the early A spectral types observed here (T$_{\rm eff} \ge 9000$~K) is
dominated by wavelengths $<1 \mu$m.  
Our input data are at 2.2 $\mu$m and longer, therefore we can not
constrain the value of the critical wavelength below 2 $\mu$m.
The grain emission
efficiency decreases at wavelengths much larger than the
grain radius and given our wavelength constraints, 
we assume that the excess is dominated by
grains with a critical wavelength of $\ge 1 \mu$m and
therefore that the absorbing efficiency is essentially constant.
The emission efficiency is assumed
to follow a power-law such that
$\epsilon_e = \epsilon_o (\lambda_o/\lambda)^q$.  This
formulation of the efficiency does not account for spectral features, but
as neither object has such features, the approximation is
appropriate.  We
investigate two values of $q$: $q=1$ which is appropriate
for absorbing dielectrics and amphorous materials
such as silicate and roughly matches the silicate population
considered in \S \ref{modeling}
and $q=2$ which is appropriate for
conductive substances such as pure graphite or crystallines and represents 
the graphite population in \S \ref{modeling}.
We then derived the temperature
of the grains as a function of radius from the star,
following \citet{bac93},
\begin{eqnarray}
T(r) = 468\ \rm{L}_{\ast}^{1/5} \lambda_o^{-1/5} r^{-2/5}\ \rm{K} \quad q=1 \\
T(r) = 685\ \rm{L}_{\ast}^{1/6} \lambda_o^{-1/3} r^{-1/3}\ \rm{K} \quad q=2 \\
\end{eqnarray}
where  L$_{\ast}$ is in L$_{\odot}$, $\lambda_o$ is in microns and $r$, the
distance to the star is in AU.  Assuming a power
law for the radial distribution, the flux in an ring is then \citep{koe98}
\begin{equation}
F(r,\lambda) = \tau_{r_o} \left( \frac{r}{r_o} \right) ^\alpha \epsilon_o \left( \frac{\lambda_o}{\lambda} \right)^q B(T) \frac{2 \pi r\, dr}{D^2}
\label{flux}
\end{equation}
where $\tau_{r_o}$ is the optical depth and $D$ is the
distance to the star from earth.  The input parameters to
our models are the inner disk radius $r_{in}$, the disk radial
extent $\Delta r$,
the optical depth, $\tau_{r_o}$, the optical depth radial exponent,
$\alpha$ and the grain characteristic wavelength $\lambda_o$.
Unless the grain composition varies with disk radius, the values of $\tau_{r_o}$ and
$\epsilon_o$ are degenerate.  We set $\epsilon_o$ = 1, thus
$\tau_{r_o}$ here represents the emission optical depth and is
only equal to the geometric optical depth if the grains have an 
albedo of 0.

For \bleo\ and \zlep\, we were unable to fit both the
near-infrared and mid-infrared excess with a single ring of dust. 
This is not surprising as the 2 $\mu$m excess is higher than
the excess flux at the shortest IRS wavelengths, requiring a 
decrease in emissivity at some intermediate radii.  The
next level of sophistication is to add a second ring of dust, with each ring
following the physical description given above.

For each object, a grid of millions of models was calculated and
compared to the input data. The results for each object are a range of
parameters consistent with the data, given all possible
values for the other parameters.  We found that some parameters were
not well constrained by the data and others were degenerate, such 
as the optical depth and the disk radial extent.  We
initially assumed that the inner and outer rings had the same radial
power-law $\alpha$ and the same characteristic grain size, $\lambda_o$.  
We were
unable to fit both the near-infrared and mid-infrared data with a two
ring model if the characteristic grain size was the same in each ring.
Fitting for two values of $\alpha$ and $\lambda_o$ within a single
grid is computationally very expensive, so we fit just the IRS data to
a single ring to constrain the values of $\alpha$ and
$\lambda_o$ for the outer ring.  For \bleo, the value of $\alpha_{\rm outer}$
is not well constrained, and we assume a value of 
-3/2 as predicted for collisionally dominated disks \citep[e.g.][]{ken05}.
 
For each star, we tried to match all the input data with the
emissivity power-laws of $\delta = 1$ or 2.  For \bleo,
the data can not be matched with $\delta = 1$, not surprising
given the flux ratio between 2 and 10 $\mu$m discussed above.  
For \zlep, models
with $\delta=1$ can match both the near-infrared and mid-infrared
data, but these models require that the outer ring extend over
20 AU, which is much larger than the extent derived by \citet{moe07}
in their imaging.  We therefore place a prior constraint of
$\Delta r_{\rm outer} < 15$ AU.  With this constraint, only
$\delta = 2$ models provide adequate fits.

\begin{figure*}[ht!]
\includegraphics[angle=270,scale=0.55]{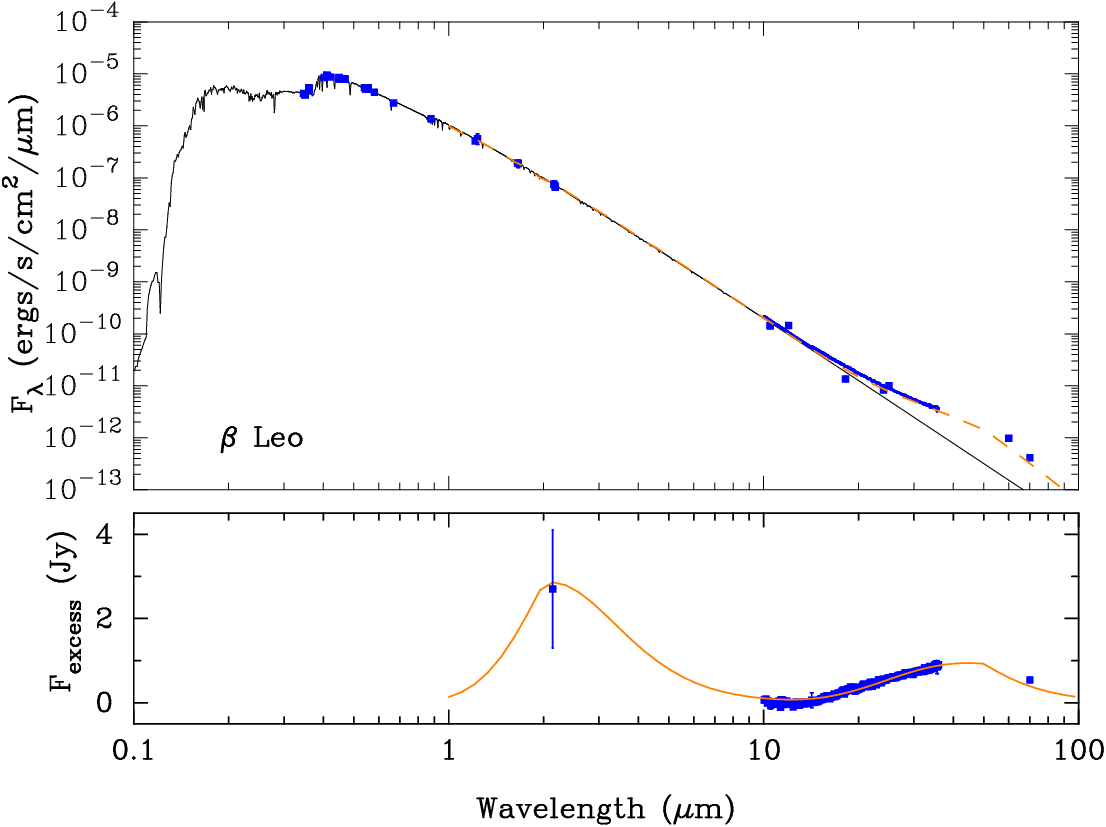}\\
\includegraphics[angle=270,scale=0.55]{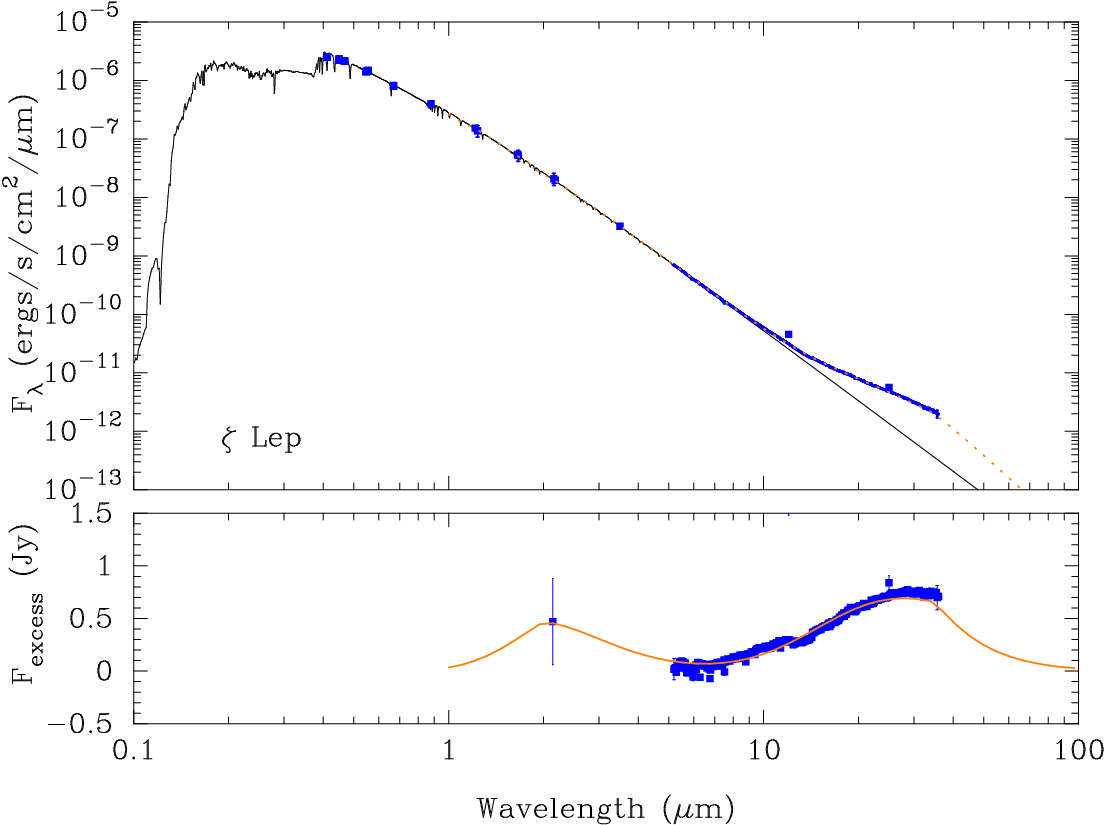}
\caption{The SED and flux excess for \bleo\ (top 2 panels) and \zlep\
(bottom 2 panels).  For each object, the top plot shows the
Kurucz-Lejeune model used for the stellar photosphere as a solid line,
with photometry from SIMBAD shown as points and the Spitzer IRS data
as a thick line with errors.  In some cases the error bars are smaller
than the points.  Our disk model is shown as a dashed line.  The
bottom plot for each object shows the disk model (solid line) with the
CHARA, IRS, and MIPS excesses.  For \bleo, the model shown has an
inner ring with $r_{\rm in}=0.13\ {\rm AU}, \Delta r = 0.3\ {\rm AU}, 
\tau_{\rm inner} =
3.8\times 10^{-3}$, $\alpha_{\rm inner}=-1.5, \lambda_{o\ \rm inner}=2 \mu$m 
and outer ring with $r_{\rm in}=13\ {\rm AU}, \Delta r = 6.2\ {\rm AU},
\tau_{\rm outer} = 3.8 \times 10^{-4}$, $\alpha_{\rm outer}$ = -1.5,
$\lambda_{o\ \rm outer}$= 50 $\mu$m.
For \zlep, the model shown has an
inner ring with $r_{\rm in}=0.16\ {\rm AU}, \Delta r = 0.05\ {\rm AU}, \tau_{\rm in} =
4.5\times 10^{-3}$, $\alpha_{\rm in}$=-2, $\lambda_{o\ \rm in}$=2 $\mu$m 
and outer ring with $r_{\rm in}=0.8\ {\rm AU}, \Delta r = 13\ {\rm AU},
\tau_{\rm outer} = 2.6 \times 10^{-4}$, $\alpha_{\rm outer}$ = -0.1,
$\lambda_{o\ \rm outer}$= 35 $\mu$m.
Both fits use a grain emissivity of $\epsilon = \epsilon_o (\lambda/\lambda_o)^{-2}$.
\label{fig:SED}} 
\end{figure*}

The range of parameter values  which  falls within a 67\%  probability
range (corresponding to $\pm 1\sigma$   for a normal distribution)  is
given for each target in  Table \ref{tab:param}.  An example model for
each  object  is   shown  with  the   SED   and  IRS  data in   Figure
\ref{fig:SED}.  For  \bleo, the  outer   ring $r_{\rm  in}$  values of
7.5-15 AU are smaller than the 19  AU found by \citet{che06} due to the
different temperature law we used,
but the inner and outer ring are clearly separated by a gap of several
AU.  For \zlep, the inner and outer ring are at similar radii ($<$ few
AU) and although the inner ring must have significantly higher opacity
to produce the near-infrared flux, it is possible to fit the data with
models  in  which the inner  and outer  ring  overlap.  Interestingly,
\citet{moe07} also required  a higher flux  ratio in their inner  ring
(2-4 AU) as compared to the outer ring (4-8 AU).

The optical depth and radial extent of the inner dust ring in these
models are degenerate parameters as the constraining data are the 2
$\mu$m flux and the lack of strong mid-infrared flux.  We have
deliberately limited these models to be optically thin, 
but we note that the near-infrared flux could also arise
from a ring with a very small radial extent which was vertically
optically thick.  
The strongest test of the radial extent for the
inner ring would be to resolve it interferometrically, which requires
observations on shorter baselines than the data presented here.  
For example, a ring around \bleo\ at
0.12 AU with a radial extent $r/4$, would have V$_{\rm ring}^2 >
0.5$ at 2 $\mu$m on baselines shorter than 13 meters and
V$_{\rm ring}^2 < 0.1$ on the 30 meter baseline we used.
High precision measurements would still be necessary given
the small flux contribution from the ring.

The relationship between the characteristic grain size and the
physical grain radius depends on the grain composition and the
distribution of sizes.  One specific example of a grain material which
could be approximated by our $\delta = 2 $ emissivity model is
graphite, which is strongly absorbing for grain radii $>$ 0.1 $\mu$m
\citep{dra84}.  To compute the physical grain size for graphite we
take $\lambda_o/a \sim 2\pi$.  The other factor is the distribution
of grain radii.  Following \citet{bac92}, one method of tying
$\lambda_o$ to the physical radii is to find the radius which divides
the grain population into two equal halves of surface area.  For a
distribution of $n(a) \propto a^{-3.5}$, this radius is 4 times the
minimum radius, $a \sim 4 a_{\rm min}$.  Putting these two factors
together, we have $a_{\rm min} \sim \lambda_o/8 \pi$.  For our upper
limit on the characteristic size in the inner ring $\lambda_o < 2
\mu$m, this corresponds to $a_{\rm min} < 0.08 \mu$m.  

Grains that small are below the nominal radiation blowout radius for
these stars.  However, radiation pressure may not completely clear all
the grains from debris disks like these.  \citet{kri00} modeled the
$\beta$ Pic debris disk, which has a similar spectral type (A6 V) and
mid-infrared excess to our targets.  In their model, small grains are
constantly created through collisions, particularly between particles
on stable orbits and those being blown out of the system.  They found
that grains smaller than a few microns were depleted compared to the
$a^{-3.5}$ distribution of a collisionally dominated disk, but that a
significant population remained.  The resulting overall grain
population could be approximated by a more shallow slope in the
distribution, for example, fitting the resulting grain radii
distribution with a single power-law between 0.1 and 100 $\mu$m,
results in $n(a) \sim a^{-2.8}$.  Thus the inner rings may contain
grains smaller than 1 $\mu$m, the nominal blowout radius.

Our model for both targets includes a much larger characteristic grain
size in the outer ring, $\lambda_o \sim 35-50 \mu$m, which corresponds
to a minimum size of $\sim 1-2 \mu$m for graphite grains.  This is
roughly the radiation blowout size.  The conclusion from our models
that the inner ring contains substantially smaller grain sizes than
the outer ring should be confirmed with more detailed grain models,
but as discussed more below, may suggest either different origins for
the grains or different dynamics.

The dust sublimation temperature of 1600~K used in these models may be
plausible for amorphous grains such as those represented by the
$\delta = 1$ model, but is higher than generally used for crystalline
grains \citep[e.g. 1250~K;][]{bau97} as represented by the $\delta =
2$ model.  A sublimation temperature of 1250~K results in much poorer
fits to the data.  At these high temperatures, micron-sized grain
lifetimes will be short; for example, \citet{lam74} found
lifetimes of less than 10$^4$ seconds for 1 $\mu$m radius grains at
1500-1600~K.  However, once the grains are very small, the grain
temperature and lifetime may increase.  In a study of grains with radii
 $< 0.01\ \mu$m heated through interactions with a single
photon, \citet{guh89} found a broad distribution of temperatures  with
excursions as high as 2800~K for graphite grains and 2050~K for
silicate grains.  \citet{guh89} calculated the sublimation rates for
these grains including a correction derived from fluctuation theory
for finite systems which decreases the sublimation rate by $\sim 10^4$.
The resulting lifetimes for grains with radii from several to
tens of Angstroms are $>10^2$ yrs.  

The combination of sublimation, radiation pressure and collisions will
result in a grain size distribution substantially more complicated
than the simple power-law often used in debris disks and assumed here.
The result of all these processes may be a population of small hot
grains which is constantly created through collisions and depleted
through sublimation and radiation pressure.  Alternatively, the presence of a
significant number of small grains may imply origin in a transient
event, as discussed in the next section.  Our formulation of the grain
temperature and emissivity efficiency does not properly represent very
small grains and more detailed models than those considered here are
necessary to determine if the temperatures and lifetimes of sub-micron
sized grains are consistent with a stable grain population which could
produce near-infrared flux observed here.  Emission from small, hot
grains has been invoked to fit the spectral energy distributions of
debris disks \citep{syl97} and the much more massive primordial disks
of Herbig Ae/Be stars \citep{nat93}.

In the models presented here, the mid-infrared excess for
\bleo\ is produced by dust grains located $r \approx 12 \pm 5$ AU from
the star. At the pixel scale of our MICHELLE/Gemini data ($0\farcs1$
pixel$^{-1}$, \S \ref{michelle}) and the distance of \bleo\ ($d =
11.1$ pc), the mid-infrared dust emission is located 
of $\sim$8 pixels from the central core of the stellar image (FWHM
$\sim 5.4$ pixels); thus, the Gemini observations of \bleo\ should
easily resolve the mid-infrared emission if the surface brightness is
high enough.  However, the Gemini observations show no evidence of
detecting the outer mid-infrared emission.  For a 5 AU ring width,
the dust emission is spread out among $\sim$400 pixels, for a surface
brightness of $\sim$0.75 mJy pixel$^{-1}$.  Given the measured
dispersion of $\sim 0.55$ mJy pixel$^{-1}$, the signal-to-noise ratio
is then only $\sim 1.4$ (per pixel).  Thus, the non-detection of the
mid-infrared excess of $\beta$~Leo is fully consistent
with our dust distribution model, even though the spatial resolution
was more than sufficient to resolve the emission.

\begin{table*}[ht!]
\begin{center}
\begin{tabular}{lll}\tableline
Model parameter & \bleo\ & \zlep\ \\ \tableline
Inner ring \\
\quad $r_{\rm in}$ (AU) & R$_{\rm sub}$ - 0.2 &  R$_{\rm sub}$ - 0.2 \\
\quad $\Delta r$ (AU) & $<$ 0.5 & $<$ 0.15 \\
\quad $\tau(2 \mu m)$ & 2-20 $\times 10^{-3}$ & 1-5 $\times 10^{-3}$ \\
\quad $\alpha$ & no constraint & $<$-1 \\
\quad $\lambda_{o}$ ($\mu$m) & $<$2  &  $<$2  \\
Outer ring \\
\quad $r_{\rm in}$ (AU) & 7 - 15 & 0.5 - 1.2\\
\quad $\Delta r$ (AU) & 3 - 10 & 11 - 15 \\
\quad $\tau(2 \mu m)$ & $1 - 4 \times 10^{-3}$ & 2-3 $\times 10^{-4}$  \\
\quad $\alpha$ & no constraint & -0.2 - 0.1 \\
\quad $\lambda_{o}$ ($\mu$m) & 35 - 70 & $>$ 30 \\ 
$\delta$  &  2 & 2 \\ \tableline
\end{tabular}
\caption{The model parameters values from fitting the
SED, IRS data and K-band excess.  The ranges given cover
a 67\% probability range.
\label{tab:param}}
\end{center}
\end{table*}

\section{Hot dust in debris disks}

Photometric and spectroscopic surveys of debris disks have revealed
other sources with dust within several AU of the star and
the frequency of these systems is a strong function of the age of the
system.  For FGK stars the frequency of debris disks emitting at
wavelengths shorter than 30$\mu$m is 9-19\% at ages less than 300 Myr,
but less than 2-4\% for stars older than 1 Gyr \citep{bei06,mey08}.
For A stars, a similar trend is seen, although the excess rate for
young A stars ($<$ 190 Myr) is even higher at 33\% \citep{su06}.  The
age estimates for our targets stars (\bleo: 50-380 Myr; \zlep: 180-490
Myr; \citet{lac99,che06}) place them in or near the age brackets for
the higher-percentage hot dust population.  Some theoretical models of
planet formation predict higher planet formation rates for A stars, as
compared to solar mass stars \citep{ken08}.  Thus, detection of hot
dust in our targets systems is not necessarily unexpected.

Other debris disks have near-infrared excesses detected through
interferometry.  Including this work, nine A and early F
stars have been observed
and, not including our tentative detection of \zlep, three (Vega,
$\zeta$ Aql and \bleo) were detected
\citep{cia01,abs06,dif07a,abs08} at the 1-2\% excess level.
Based on limited radial velocity data, \citet{abs08} suggest that their detected excess toward
$\zeta$ Aql may arise from a close M star companion and based on
a reanalysis of the Spitzer data, they also find that there is
no mid-infrared excess from this source. 
Only one lower mass star, $\tau$
Ceti, has an observed near-infrared excess \citep{dif07a}, although the
observations of FGK stars are currently limited by the
sensitivity of the instruments.  The near-infrared flux levels of the
detected sources are all at the few percent level, although this may
represent the brightest examples of a population of disks with hot
grains, as the limits on the non-detected sources are not
substantially lower, e.g. \citet{dif07a} set an upper limit of 0.6\%
for the near-infrared emission from $\epsilon$ Eri.

The dust distribution inferred for Vega by \citet{abs06} is
somewhat similar to \bleo\ in that they modeled the
inner dust with an inner radius of 0.17 to 0.3 AU with
a very steep ($n(r) \propto r^{-4}$) radial power law, which is similar
to a ring.  The inner dust mass inferred was $8 \times 10^{-8}$
M$_{\oplus}$ with a dust luminosity of $5 \times 10^{-4}$~L$_{\odot}$.

Although the number of debris disk stars surveyed is still relatively
small, there are three detections (one marginally significant) 
of near-infrared excess
for which a population of hot dust is the most likely explanation, and
here we consider if they represent a stable dust population
produced by collisions of larger bodies orbiting close to the star or
are the result of a transient event.  The production of small, hot
dust grains from the break-up of a comet or asteroid has been invoked
in other cases, such as HD 69830 \citep{bei05}.  Using a density of 3
gm cm$^{-3}$, a mass of $ 5 \times 10^{-9}$~M$_{\oplus}$ can be generated from
the break-up of a single body with a 10 km radius.

To evaluate the likelihood of so many transient events in A stars, we
used the model of \citet{wya07} for the steady-state evolution of
collision-dominated debris disks, in which they derive a maximum
possible dust luminosity as a function of age. In examining the
properties of 46 known A-star debris disks, they found only 4 stars
with a dust luminosity significantly higher than this maximum.  A dust
luminosity well above the steady-state maximum suggests either a
transient origin for the dust or unusual properties for the
planetesimal belt.  The mid-infrared excess of \zlep\ is above this
threshold, while the mid-infrared excess of \bleo\ is below.  We use
the same calculation to evaluate the inner dust.  For simplicity, we
take the maximum steady-state flux ($f_{\rm max}$) derived by
\citet{wya07} for the outer planetesimal ring and scale that value to
the inner ring radius using their derived relationship $f_{\rm max}
\propto r^{7/3}$.  This ignores any difference in grain properties
between the disks but is acceptable for an order of magnitude calculation.  For
both \bleo\ and \zlep\, the near-infrared dust luminosity ($f$)
compared to the maximum allowed is $f/f_{\rm max} \sim 10^6$,
obviously above the threshold of 10 set by \citet{wya07} for anomalous
systems.  We also calculated this quantity for Vega, which for the
inner dust also has a value of $f/f_{\rm max} \sim 10^6$.  In this
model of collisionally-dominated disks, all three of these objects are
orders of magnitude higher than the expected steady-state flux,
suggesting a transient event as the most likely origin.  The finding
from our simple dust model of small dust within the inner ring, but
not the outer, mid-infrared producing ring, may also favor a transient origin
for the near-infrared producing dust given the issues of
dust lifetime to radiation pressure and sublimation.  A recent planetesimal collision
or comet passing would drastically change the dust radii distribution
and dynamics.

Although more objects should be sampled to come to a stronger
conclusion, it is suggestive that the near-infrared excess in 
these objects arises from a recent collision or cometary passing event.
The near-infrared excess is an ideal probe of hot, small grains
in these systems,
as dust much closer to the star will sublimate.  
An observational test of the hypothesis that the near-infrared
flux arises from emission from grains near the sublimation radius
is to make observations at other wavelengths, particularly
H (1.6 $\mu$m) and L (3.5$\mu$m) bands, to probe the wavelength
dependence of the excess.  If the flux is dominated by emission,
the peak will be near K and L, while if it is scattering, the excess at H
will be much higher and the excess at L much lower.

\section{Conclusions}

We have presented near-infrared interferometry observations of two A
stars, \bleo\ and \zlep, which were known to have mid-infrared excess
emission from a debris disk.  A near-infrared excess of 1-2\% was
detected, although the detection for \zlep\ should be confirmed.  The
interferometer observations do not spatially resolve the emission
distribution, but place a maximum on the radial extent through the field of
view, and in conjunction with the spectral energy distribution, the
spatial distribution of dust can be constrained. Both objects
can be modeled as having a thin ring of dust grains at or
near the sublimation radius in addition to the previously
known mid-infrared emitting belt.  

Although the models presented here are not a unique fit to the data,
particularly with respect to the grain population, we can place strong
constraints on the dust composition and morphology.  Both objects
require small, non-silicate grains to be consistent with the
near-infrared and mid-infrared excesses.  The minimum grain size
required ($\sim 0.1 \mu$m) is an order of magnitude smaller than the
nominal radiation pressure blowout radius for spherical grains and
requires a high production rate of small grains if some depletion does
occur due to radiation pressure and sublimation.  For \bleo, the near-infrared excess
can not arise from dust generated by the planetesimal belt which
produces the mid-infrared excess.  For \zlep, the most likely model
using simple geometric distributions is also two separate rings, but
it is possible that the larger bodies producing the inner dust may
form a continuous extent with the mid-infrared planetesimal belt,
although not with a simple, single power-law radial distribution.  The
luminosity of the inner dust is exceptionally high in comparison to
steady-state evolutionary models of collisionally-dominated debris
disks, suggesting origin in a transient event, such as the break-up of
a comet or asteroid near the star.

These observations are limited by the sensitivity of the current
instruments, but with improvements in near-infrared interferometry and
other techniques, such as nulling interferometry and adaptive optics
with coronagraphy, it should be possible to survey many more stars to
determine the population with hot dust.

\acknowledgments

We thank the CHARA staff, particularly P.J. Goldfinger, for their
excellent help in obtaining the data and the FLUOR team for support of
the instrument.  We thank Christine Chen for kindly providing IRS
spectra, Karl Stapelfeldt for the MIPS measurement and Mark Wyatt,
Scott Kenyon and Hal Levison for helpful discussions.  The anonymous
referee made several helpful suggestions to improve the paper.  This
work was performed at the Michelson Science Center, Caltech and made
use of the SIMBAD database, operated at CDS, Strasbourg, France and
the NASA Star and Exoplanet Database (NStED) at the Infrared
Processing and Analysis Center. NStED is jointly funded by the
National Aeronautics and Space Administration (NASA) via Research
Opportunities in Space Sciences grant 2003 TPF-FS, and by NASA's
Michelson Science Center.

{\it Facilities:} \facility{CHARA,GEMINI}.


\end{document}